\newcommand{\CLASSINPUTinnersidemargin}{1in}
\newcommand{\CLASSINPUToutersidemargin}{1in}
\newcommand{\CLASSINPUTtoptextmargin}{1in}
\newcommand{\CLASSINPUTbottomtextmargin}{1in}
\documentclass[journal,12pt,onecolumn,draftcls]{IEEEtran}
\usepackage{stfloats}
\ifCLASSINFOpdf
\usepackage{graphicx}
 \usepackage{epsfig}
  \usepackage{epstopdf}
 \graphicspath{{F:\Latex\latex_hamed\IEEE_Communications_Letter\AEN\IEEEtran}}
   \DeclareGraphicsExtensions{.eps,.pdf,.jpeg,.png,.tif}

\else
\fi

\usepackage[cmex10]{amsmath}
\usepackage{latexsym}
\usepackage{amssymb}

\newcounter{MYtempeqncnt}
\begin{document}
%
\title{Comments on ``Achievable Rates in Cognitive Radio Channels'' }

\author{Mostafa~Monemizadeh 
\thanks{Manuscript received September, 2015.}%
\thanks{The author is with the Department of Electrical Engineering, Ferdowsi University of Mashhad, Iran (e-mail: mostafamonemizadeh@gmail.com}}%


\maketitle

\begin{abstract}
In a premier paper on the information-theoretic analysis of a two-user cognitive interference channel (CIC) \cite{DMT:CIC}, Devroye \textit{et al.} presented an achievable rate region for the two-user discrete memoryless CIC. The coding scheme proposed by Devroye \textit{et al.} is correct but unfortunately some rate-terms in the derived achievable rate region are incorrect (in fact incomplete) because of occurring some mistakes in decoding and analysis of error probability. We correct and complete the wrong rate-terms and thereby show that the corrected achievable rate region includes the rate region presented in \cite{DMT:CIC}.
\end{abstract}


\IEEEpeerreviewmaketitle

\section{Introduction}
A two-user genie-aided cognitive interference channel (CIC) is a two-user interference channel (IC) in which one of the transmitters (termed the secondary transmitter, here $TX2$) knows the other transmitter's message (termed the primary transmitter, here $TX1$) noncausally (i.e., by a genie). Devroye, Mitran, and Tarokh (DMT) in their premier paper, titled ``Achievable Rates in Cognitive Radio Channels,'' derived an achievable rate region for the discrete memoryless CIC ([\ref{DMT:CIC}, Th. 1]). We observe that the coding scheme proposed in \cite{DMT:CIC} is correct but unfortunately the derived achievable rate region is incorrect because of occurring some mistakes in decoding and analysis of error probability. 
 
We first intuitively show that some rate-terms in the DMT rate region seem to be incorrect (in fact, they are incomplete). Then, we correct the DMT achievable rate region and thereby show that the corrected achievable rate region includes the DMT rate region given in \cite{DMT:CIC}. 

\section{Preliminaries}
The two-user discrete memoryless CIC (DM-CIC), denoted by $\{\mathcal{X}_1\times \mathcal{X}_2,p(y_1,y_2 |x_1,x_2),\mathcal{Y}_1\times \mathcal{Y}_2\},$ consists of four finite alphabets $\mathcal{X}_1,\mathcal{X}_2,\mathcal{Y}_1,\mathcal{Y}_2,$ and a collection of conditional probability mass functions $p(y_1,y_2 |x_1,x_2)$ on $\mathcal{Y}_1\times \mathcal{Y}_2$. The channel is memoryless in the sense that $p(y_1^n,y_2^n | x_1^n,x_2^n )=\prod _{t=1}^n p(y_{1,t},y_{2,t} |x_{1,t},x_{2,t})$.

In this channel transmitter $i,~i \in \{1,2\}$, wants to send a message $M_i$, uniformly distributed on $\mathcal{M}_i=\left\{1,\cdots,2^{nR_i}\right\}$, to its respective receiver. The primary transmitter $TX1$ generates the codeword $x_1^n$ as $f_1: \mathcal{M}_1\rightarrow \mathcal{X}_1^n$, and the secondary transmitter $TX2$, being non-causally aware of the primary message, generates the codeword $x_2^n$ as $f_2: \mathcal{M}_1\times \mathcal{M}_2\rightarrow \mathcal{X}_2^n$. The decoding function $g_i (\cdot)$ is given by $g_i: \mathcal{Y}_i^n\rightarrow \mathcal{M}_i$.

A pair $(R_1,R_2)$ of non-negative real values is called an achievable rate for the DM-CIC if for any given $0<\epsilon<1$ and for any sufficiently large $n$, there exists a sequence of encoding functions $f_1 (\cdot), f_2 (\cdot)$, and a sequence of decoding functions $g_1 (\cdot), g_2 (\cdot)$, such that 
\begin{IEEEeqnarray}{l}
P_e^{(n)}=P_r\Big\{ g_1(y_1^n) \neq m_1 ~~\mathrm{or}~~ g_2(y_2^n) \neq m_2~|~ (m_1,m_2) ~\mathrm{sent} \Big\} \leq \epsilon \nonumber
\end{IEEEeqnarray}
where $P_e^{(n)}$ is the average probability of error. The closure of the set of all achievable rate pairs $(R_1,R_2)$ is called the capacity region.

In \cite{DMT:CIC}, Devroye \textit{et al.}, by using rate splitting, divided each message $M_i$, $i \in \{1,2\}$, into two independent sub-messages: 
\begin{itemize}
\item [(i)] common sub-message $M_{ic}$ at rate $R_{ic}$ (to be sent from $TXi \rightarrow RX1,RX2$), 
\item[(ii)] private sub-message $M_{ip}$ at rate $R_{ip}$ (to be sent from $TXi \rightarrow RXi$), 
\end{itemize}\vspace{5pt}
such that $R_i=R_{ic}+R_{ip}$.
In this paper, auxiliary random variables (RVs) $U_{ic}$ and $U_{ip},~i \in \{1,2\},$ represent the sub-messages $M_{ic}$ and $M_{ip}$, respectively. Moreover, RV $Q$ is time sharing RV which is independent of all other RVs. 

We now present the DMT achievable rate region for the two-user genie-aided DM-CIC. \vspace{5pt}

\textit{Theorem 1 [\ref{DMT:CIC}, Th. 1]:} Let ${\mathcal P}^{DMT}$ be the set of all joint distributions $p(\cdot)$ that factor as 
\begin{IEEEeqnarray}{l}
p\left(q,u_{1c},u_{1p},u_{2c},u_{2p},x_1,x_2\right)= \nonumber \\
~~~p(q)p(u_{1c}|q)p(u_{1p}|q)p(x_1|q,u_{1c},u_{1p})p(u_{2c}|q,u_{1c},u_{1p})p(u_{2p}|q,u_{1c},u_{1p})p(x_2|q,u_{2c},u_{2p}).~~
\label{pdf_DMT} 
\end{IEEEeqnarray}
For any $p(\cdot)\in{\mathcal P}^{DMT}$, let ${\mathcal R}^{DMT}(p)$ be the set of all quadruples $(R_{1p},R_{1c},R_{2c},R_{2p})$ of non-negative real numbers such that there exist non-negative real $(R_{2c}^{'},R_{2p}^{'})$ satisfying
{\small
\begin{align}
\label{DMT}
R_{2c}^{'} & \geq I(U_{2c};U_{1p},U_{1c}|Q) \tag{2.1}\\ 
R_{2p}^{'} & \geq I(U_{2p};U_{1p},U_{1c}|Q) \tag{2.2}\\ 
R_{1p} & \le I(Y_1;U_{1p}|U_{1c},U_{2c},Q)+I(U_{2c};U_{1p}|U_{1c},Q) \tag{2.3}\\ 
R_{1c} & \le I(Y_1;U_{1c}|U_{1p},U_{2c},Q)+I(U_{2c};U_{1c}|U_{1p},Q) \tag{2.4}\\ 
R_{2c}+R_{2c}^{'} & \le I(Y_1;U_{2c}|U_{1p},U_{1c},Q) +I(U_{1p},U_{1c};U_{2c}|Q)\tag{2.5}\\ 
R_{1p}+R_{1c} & \le I(Y_1;U_{1p},U_{1c}|U_{2c},Q)+I(U_{2c};U_{1p},U_{1c}|Q) \tag{2.6}\\ 
R_{1p}+R_{2c}+R_{2c}^{'} & \le I(Y_1;U_{1p},U_{2c}|U_{1c},Q)+I(U_{2c};U_{1c}|U_{1p},Q) \tag{2.7}\\ 
R_{1c}+R_{2c}+R_{2c}^{'} & \le I(Y_1;U_{1c},U_{2c}|U_{1p},Q)+I(U_{2c};U_{1p}|U_{1c},Q) \tag{2.8}\\ 
R_{1p}+R_{1c}+R_{2c}+R_{2c}^{'} & \le I(Y_1;U_{1p},U_{1c},U_{2c}|Q) \tag{2.9}\\
R_{2p}+R_{2p}^{'} & \le I(Y_2;U_{2p}|U_{1c},U_{2c},Q)+I(U_{1c},U_{2c};U_{2p}|Q) \tag{2.10}\\ 
R_{2c}+R_{2c}^{'} & \le I(Y_2;U_{2c}|U_{2p},U_{1c},Q)+I(U_{1c},U_{2p};U_{2c}|Q) \tag{2.11}\\ 
R_{1c} & \le I(Y_2;U_{1c}|U_{2p},U_{2c},Q)+I(U_{2p},U_{2c};U_{1c}|Q) \tag{2.12}\\ 
R_{2p}+R_{2p}^{'}+R_{2c}+R_{2c}^{'} & \le I(Y_2;U_{2p},U_{2c}|U_{1c},Q)+I(U_{1c};U_{2p},U_{2c}|Q)~~~~~~~~~~~~~~~~~~ \tag{2.13}\\ 
R_{2p}+R_{2p}^{'}+R_{1c} & \le I(Y_2;U_{2p},U_{1c}|U_{2c},Q)+(U_{2c};U_{2p},U_{1c}|Q) \tag{2.14}\\ 
R_{2c}+R_{2c}^{'}+R_{1c} & \le I(Y_2;U_{2c},U_{1c}|U_{2p},Q)+I(U_{2p};U_{2c},U_{1c}|Q) \tag{2.15}\\ 
R_{2p}+R_{2p}^{'}+R_{2c}+R_{2c}^{'}+R_{1c} & \le I(Y_2;U_{2p},U_{2c},U_{1c}|Q) \tag{2.16}
\end{align}
}then  
{\begin{itemize}
\item [(i)] ${\mathcal R}^{DMT}\stackrel{\textrm{def}}{=}\bigcup_{p(\cdot)\in{\mathcal P}^{DMT}}{\mathcal R}^{DMT}(p)$ is an achievable rate region for the genie-aided DM-CIC in terms of $(R_{1p},R_{1c},R_{2c},R_{2p})$,
\item[(ii)] ${\mathcal R}^{DMT}_{imp}\stackrel{\textrm{def}}{=}\bigcup_{p(\cdot)\in{\mathcal P}^{DMT}}{\mathcal R}^{DMT}_{imp}(p)$ is the implicit description of the DMT achievable rate region where ${\mathcal R}^{DMT}_{imp}(p)$ is the set of all pairs $(R_1,R_2)$ of non-negative real numbers such that $R_1=R_{1p}+R_{1c}$ and $R_2=R_{2p}+R_{2c}$ for some $(R_{1p},R_{1c},R_{2c},R_{2p})\in{\mathcal R}^{DMT}(p)$.
\end{itemize}
}

\setcounter{MYtempeqncnt}{\value{equation}}
\setcounter{equation}{2}

\section{Main Results}
We first saw that the rate-terms (2.7)--(2.9) and (2.13)--(2.16) of ${\mathcal R}^{DMT}(p)$ intuitively seem to be incomplete because of not utilizing some dependencies among RVs. For example, in (2.7) we have the main term $I(Y_1;U_{1p},U_{2c}|U_{1c},Q)$. By considering the coding exploited in \cite{DMT:CIC} and ${\mathcal P}^{DMT}$ and also since in the main term $I(Y_1;U_{1p},U_{2c}|U_{1c},Q)$, RV $U_{1c}$ is known (or given) and RVs $U_{1p},U_{2c}$ are unknown, we expect that the dependency between known RV $U_{1c}$ and unknown RVs $(U_{1p},U_{2c})$ as well as the dependency between unknown RVs $U_{1p}$ and $U_{2c}$ help communication and boost the rates. As we observe in (2.7), the term $I(U_{1p},U_{2c};U_{1c}|Q)=I(U_{2c};U_{1c}|U_{1p},Q)$ is added to the main term but unfortunately, the term $I(U_{1p};U_{2c}|Q)$ is not. Similarly, in (2.8) the dependency between $U_{1c}$ and $U_{2c}$, in (2.9) the dependency between $(U_{1p},U_{1c})$ and $U_{2c}$, in (2.13) the dependency between $U_{2p}$ and $U_{2c}$, in (2.14) the dependency between $U_{2p}$ and $U_{1c}$, in (2.15) the dependency between $U_{1c}$ and $U_{2c}$, and in (2.16) the dependencies among $U_{1c}$, $U_{2p}$ and $U_{2c}$ can help communication and boost the rates, while they are overlooked in ${\mathcal R}^{DMT}(p)$. In this section, we present the corrected version of the DMT rate region that utilizes the aforementioned dependencies among auxiliary RVs in boosting the rates. \vspace{10pt}    

\textit{Theorem 2 [corrected Th. 1]:} For any $p(\cdot)\in{\mathcal P}^{DMT}$, let ${\mathcal R}^{Co-DMT}(p)$ be the set of all quadruples $(R_{1p},R_{1c},R_{2c},R_{2p})$ of non-negative real numbers such that there exist non-negative real $(R_{2c}^{'},R_{2p}^{'})$ satisfying
{\small
\begin{align}
\label{corrected_DMT}
R_{2c}^{'} & \geq I(U_{2c};U_{1p},U_{1c}|Q) \tag{3.1}\\ 
R_{2p}^{'} & \geq I(U_{2p};U_{1p},U_{1c}|Q) \tag{3.2}\\ 
R_{1p} & \le I(Y_1;U_{1p}|U_{1c},U_{2c},Q)+I(U_{2c};U_{1p}|U_{1c},Q) \tag{3.3}\\ 
R_{1c} & \le I(Y_1;U_{1c}|U_{1p},U_{2c},Q)+I(U_{2c};U_{1c}|U_{1p},Q) \tag{3.4}\\ 
R_{2c}+R_{2c}^{'} & \le I(Y_1;U_{2c}|U_{1p},U_{1c},Q) +I(U_{1p},U_{1c};U_{2c}|Q)\tag{3.5}\\ 
R_{1p}+R_{1c} & \le I(Y_1;U_{1p},U_{1c}|U_{2c},Q)+I(U_{2c};U_{1p},U_{1c}|Q) \tag{3.6}\\ 
R_{1p}+R_{2c}+R_{2c}^{'} & \le I(Y_1;U_{1p},U_{2c}|U_{1c},Q)+I(U_{2c};U_{1c}|U_{1p},Q)+\mbox{\boldmath $I(U_{2c};U_{1p}|Q)$} \tag{3.7}\\
R_{1c}+R_{2c}+R_{2c}^{'} & \le I(Y_1;U_{1c},U_{2c}|U_{1p},Q)+I(U_{2c};U_{1p}|U_{1c},Q)+\mbox{\boldmath $I(U_{2c};U_{1c}|Q)$} \tag{3.8}\\ 
R_{1p}+R_{1c}+R_{2c}+R_{2c}^{'} & \le I(Y_1;U_{1p},U_{1c},U_{2c}|Q)+\mbox{\boldmath $I(U_{2c};U_{1p},U_{1c}|Q)$} \tag{3.9}\\
R_{2p}+R_{2p}^{'} & \le I(Y_2;U_{2p}|U_{2c},U_{1c},Q)+I(U_{2c},U_{1c};U_{2p}|Q) \tag{3.10}\\ 
R_{2c}+R_{2c}^{'} & \le I(Y_2;U_{2c}|U_{2p},U_{1c},Q)+I(U_{2p},U_{1c};U_{2c}|Q) \tag{3.11}\\ 
R_{1c} & \le I(Y_2;U_{1c}|U_{2p},U_{2c},Q)+I(U_{2p},U_{2c};U_{1c}|Q) \tag{3.12}\\ 
R_{2p}+R_{2p}^{'}+R_{2c}+R_{2c}^{'} & \le I(Y_2;U_{2p},U_{2c}|U_{1c},Q)+I(U_{1c};U_{2p},U_{2c}|Q)+\mbox{\boldmath $I(U_{2p};U_{2c}|Q)$} \tag{3.13}\\ 
R_{2p}+R_{2p}^{'}+R_{1c} & \le I(Y_2;U_{2p},U_{1c}|U_{2c},Q)+(U_{2c};U_{2p},U_{1c}|Q)+\mbox{\boldmath $I(U_{2p};U_{1c}|Q)$} \tag{3.14}\\ 
R_{2c}+R_{2c}^{'}+R_{1c} & \le I(Y_2;U_{2c},U_{1c}|U_{2p},Q)+I(U_{2p};U_{2c},U_{1c}|Q)+\mbox{\boldmath $I(U_{2c};U_{1c}|Q)$} \tag{3.15}\\ 
R_{2p}+R_{2p}^{'}+R_{2c}+R_{2c}^{'}+R_{1c} & \le I(Y_2;U_{2p},U_{2c},U_{1c}|Q)+\mbox{\boldmath $I(U_{2p},U_{2c};U_{1c}|Q)$}+\mbox{\boldmath $I(U_{2p};U_{2c}|Q)$} \tag{3.16}
\end{align}
}then
{\begin{itemize}
\item [(i)] ${\mathcal R}^{Co-DMT}\stackrel{\textrm{def}}{=}\bigcup_{p(\cdot)\in{\mathcal P}^{DMT}}{\mathcal R}^{Co-DMT}(p)$ is an achievable rate region for the genie-aided DM-CIC in terms of $(R_{1p},R_{1c},R_{2c},R_{2p})$,
\item[(ii)] ${\mathcal R}^{Co-DMT}_{imp}\stackrel{\textrm{def}}{=}\bigcup_{p(\cdot)\in{\mathcal P}^{DMT}}{\mathcal R}^{Co-DMT}_{imp}(p)$ is the implicit description of the corrected DMT achievable rate region where ${\mathcal R}^{Co-DMT}_{imp}(p)$ is the set of all pairs $(R_1,R_2)$ of non-negative real numbers such that $R_1=R_{1p}+R_{1c}$ and $R_2=R_{2p}+R_{2c}$ for some $(R_{1p},R_{1c},R_{2c},R_{2p})\in{\mathcal R}^{Co-DMT}(p)$.
\end{itemize}
}
\setcounter{MYtempeqncnt}{\value{equation}}
\setcounter{equation}{3}

\textit{Remark 1:} Because of some added mutual information terms (boldface terms in Theorem 2), the corrected DMT rate region includes the previous incomplete one. 
\vspace{15pt}

\textit{Proof of Theorem 2:}

\textit{Codebook generation: }Fix a joint distribution $p(\cdot)\in {\mathcal P}^{DMT}$ factored as \eqref{pdf_DMT}. Generate a sequence $q^n\sim\prod^n_{t=1}{p\left(q_{t}\right)}$. Note that by considering
\begin{IEEEeqnarray}{l}
p\left(u_{2c}|q\right)=\sum_{u_{1c},u_{1p}~\in~\mathcal{U}_{1c},\mathcal{U}_{1p}}p\left(u_{2c}|u_{1c},u_{1p},q\right)p\left(u_{1c}|q\right)p\left(u_{1p}|q\right)~~~~~~~~~~~~~~~~~~~~~~~~~~~~~~~~~\\
p\left(u_{2p}|q\right)=\sum_{u_{1c},u_{1p}~\in~\mathcal{U}_{1c},\mathcal{U}_{1p}}p\left(u_{2p}|u_{1c},u_{1p},q\right)p\left(u_{1c}|q\right)p\left(u_{1p}|q\right)
\label{Bin.distrib}
\end{IEEEeqnarray}
the generation of the codewords $u^n_{2c}$ and $u^n_{2p}$ can be performed independently of $u^n_{1c}$ and $u^n_{1p}$ by using binning scheme. In other words, the codebook is generated according to the distribution  
\begin{IEEEeqnarray}{l}
p(q)p(u_{1c}|q)p(u_{1p}|q)p(x_1|q,u_{1c},u_{1p})p(u_{2c}|q)p(u_{2p}|q)p(x_2|q,u_{2c},u_{2p}).~~
\label{pdf_DMT_codebook} 
\end{IEEEeqnarray}
To do so,
\begin{itemize}
\item [(1)] generate $2^{nR_{1c}}$ independent and identically distributed (i.i.d.) \textit{n}-sequences $u^n_{1c}(m_{1c}),$ $m_{1c}\in \left\{1,2, \cdots ,2^{nR_{1c}}\right\}$, each according to $\prod^n_{t=1}{p\left(u_{1c,t}|q_t\right)}$; 
\item [(2)] generate $2^{nR_{1p}}$ i.i.d. \textit{n}-sequences $u^n_{1p}(m_{1p}),$ $m_{1p}\in \left\{1,2, \cdots ,2^{nR_{1p}}\right\}$, each according to $\prod^n_{t=1}{p\left(u_{1p,t}|q_t\right)}$; 
\item [(3)] generate $2^{n(R_{2c}+R^{'}_{2c})}$ i.i.d. \textit{n}-sequences $u^n_{2c}(m_{2c},l_{2c}),$ $m_{2c}\in \left\{1,2, \cdots ,2^{nR_{2c}}\right\}$ and $ l_{2c}\in \left\{1,2, \cdots ,2^{nR^{'}_{2c}}\right\}$, each according to $\prod^n_{t=1}{p\left(u_{2c,t}|q_t\right)}$; (i.e., $2^{nR_{2c}}$ bins and $2^{nR^{'}_{2c}}$ sequences in each bin)
\item [(4)] generate $2^{n(R_{2p}+R^{'}_{2p})}$ i.i.d. \textit{n}-sequences $u^n_{2p}(m_{2p},l_{2p}),$ $m_{2p}\in \left\{1,2, \cdots ,2^{nR_{2p}}\right\}$ and $l_{2p}\in \left\{1,2, \cdots ,2^{nR^{'}_{2p}}\right\},$ each according to $\prod^n_{t=1}{p\left(u_{2p,t}|q_t\right)}$; (i.e., $2^{nR_{2p}}$ bins and $2^{nR^{'}_{2p}}$ sequences in each bin)
\end{itemize}
The aim is to send a four dimensional message consisting of four sub-messages as
\begin{IEEEeqnarray}{l}
(m_{1p},m_{1c},m_{2c},m_{2p})\in \left\{1, \cdots ,2^{nR_{1p}}\right\} \times \left\{1, \cdots ,2^{nR_{1c}}\right\}\times \left\{1, \cdots ,2^{nR_{2c}}\right\}\times \left\{1, \cdots ,2^{nR_{2p}}\right\} \nonumber
\end{IEEEeqnarray}
Note that $m_{1p}$ and $m_{1c}$ are message indices and $m_{2p}$ and $m_{2c}$ are bin indices.

\textit{Encoding: }The primary transmitter $TX1$ to send $(m_{1p},m_{1c})$, first looks up the sequences $u^n_{1p}(m_{1p})$ and $u^n_{1c}(m_{1c})$, then generates $x^n_1$ i.i.d. according to $\prod^n_{t=1}{p\left(x_{1,t}|u_{1p,t}(m_{1p}),u_{1c,t}(m_{1c}),q_t\right)}$ and sends it.  

The cognitive transmitter $TX2$, being non-causally aware of $u^n_{1p}(m_{1p})$ and $u^n_{1c}(m_{1c})$, to send $m_{2p}$ and $m_{2c}$, first looks for indices $l_{2p}$ and $l_{2c}$ in bins $m_{2p}$ and $m_{2c}$, respectively, such that 
\begin{IEEEeqnarray}{l}
\left(q^n,u^n_{1p}(m_{1p}),u^n_{1c}(m_{1c}),u^n_{2p}(m_{2p},l_{2p})\right)\in A^{(n)}_{\varepsilon }\left(Q,U_{1p},U_{1c},U_{2p}\right) \IEEEnonumber\\
\left(q^n,u^n_{1p}(m_{1p}),u^n_{1c}(m_{1c}),u^n_{2c}(m_{2c},l_{2c})\right)\in A^{(n)}_{\varepsilon }\left(Q,U_{1p},U_{1c},U_{2c}\right) \IEEEnonumber
\label{typical1}
\end{IEEEeqnarray}
then generates $x^n_2$ i.i.d. according to $\prod^n_{t=1}{p\left(x_{2,t}|u_{2p,t}(m_{2p},l_{2p}),u_{2c,t}(m_{2c},l_{2c}),q_t\right)}$ and sends it.  

\textit{Decoding: }Upon receiving $y_1^n$, receiver $RX1$ looks for a unique triplet $(m_{1p},m_{1c},m_{2c})$ and some $l_{2c}$ such that
\begin{IEEEeqnarray}{l}
\left(q^n,u^n_{1p}(m_{1p}),u^n_{1c}(m_{1c}),u^n_{2c}(m_{2c},l_{2c}),y^n_1\right)\in A^{(n)}_{\varepsilon }\left(Q,U_{1p},U_{1c},U_{2c},Y_1\right) \IEEEnonumber
\label{typical2}
\end{IEEEeqnarray}

Upon receiving $y_2^n$, receiver $RX2$ looks for a unique triplet $(m_{2p},m_{2c},m_{1c})$ and some $(l_{2p},l_{2c})$ such that
\begin{IEEEeqnarray}{l}
\left(q^n,u^n_{2p}(m_{2p},l_{2p}),u^n_{2c}(m_{2c},l_{2c}),u^n_{1c}(m_{1c}),y^n_2\right)\in A^{(n)}_{\varepsilon }\left(Q,U_{2p},U_{2c},U_{1c},Y_2\right) \IEEEnonumber
\label{typical3}
\end{IEEEeqnarray}

\textit{Error analysis: }Assume without loss of generality that the message $\left(m_{1p},m_{1c},m_{2c},m_{2p}\right)=\left(1,1,1,1\right)$ is sent. We first do the encoding error analysis. The encoding error events at encoder 2 are  
\begin{IEEEeqnarray}{l}
E^{enc2}_1=\Big\{\left(q^n,u^n_{1p}(1),u^n_{1c}(1),u^n_{2c}(1,l_{2c})\right)\notin A^{(n)}_{\varepsilon }\left(Q,U_{1p},U_{1c},U_{2c}\right) ~{\rm for~all}~l_{2c}\in \{1, \cdots ,2^{nR^{'}_{2c}}\}\Big\} \IEEEnonumber\\
E^{enc2}_2=\Big\{\left(q^n,u^n_{1p}(1),u^n_{1c}(1),u^n_{2p}(1,l_{2p})\right)\notin A^{(n)}_{\varepsilon }\left(Q,U_{1p},U_{1c},U_{2p}\right) ~{\rm for~all}~l_{2p}\in \{1, \cdots ,2^{nR^{'}_{2p}}\}\Big\} \IEEEnonumber
\label{Enc.error}
\end{IEEEeqnarray} 
As shown in \cite{DMT:CIC}, the probabilities of encoding error events $E^{enc2}_1$ and $E^{enc2}_2$ go to zero as $n\to \infty $ if the following binning conditions hold: 
\begin{IEEEeqnarray}{l}
R_{2c}^{'} \geq I(U_{2c};U_{1p},U_{1c}|Q) \\ 
R_{2p}^{'} \geq I(U_{2p};U_{1p},U_{1c}|Q)    
\label{Bin.con}  
\end{IEEEeqnarray}
Let $L^*_{2c}$ and $L^*_{2p}$ denote the right Gel’fand-Pinsker coding indices (\cite{Gelfand:Pinsker}) chosen by encoder 2, i.e.,
\begin{IEEEeqnarray}{l}
\left(q^n,u^n_{1p}(1),u^n_{1c}(1),u^n_{2c}(1,L^*_{2c})\right)\in A^{(n)}_{\varepsilon }\left(Q,U_{1p},U_{1c},U_{2c}\right) \\
\left(q^n,u^n_{1p}(1),u^n_{1c}(1),u^n_{2p}(1,L^*_{2p})\right)\in A^{(n)}_{\varepsilon }\left(Q,U_{1p},U_{1c},U_{2p}\right)
\label{right.GP.index}
\end{IEEEeqnarray} 
By assuming successful encoding with no errors, we now do the decoding error analysis at decoders. The determining error events at decoder 1 are 
\begin{IEEEeqnarray}{l}
E^{dec1}_{1}=\big\{\left(q^n,u^n_{1p}(m_{1p}),u^n_{1c}(1),u^n_{2c}(1,L^*_{2c}),y^n_1\right)\in A^{(n)}_{\varepsilon 1} ~~ {\rm  for}~ m_{1p} \ne 1 \big\}\IEEEnonumber\\
E^{dec1}_{2}=\big\{\left(q^n,u^n_{1p}(1),u^n_{1c}(m_{1c}),u^n_{2c}(1,L^*_{2c}),y^n_1\right)\in A^{(n)}_{\varepsilon 1} ~~ {\rm  for}~ m_{1c} \ne 1 \big\}\IEEEnonumber\\
E^{dec1}_{3}=\big\{\left(q^n,u^n_{1p}(1),u^n_{1c}(1),u^n_{2c}(m_{2c},l_{2c}),y^n_1\right)\in A^{(n)}_{\varepsilon 1} ~~ {\rm  for}~ m_{2c} \ne 1 ~{\rm  and }~l_{2c} \ne L^*_{2c} \big\}\IEEEnonumber\\
E^{dec1}_{4}=\big\{\left(q^n,u^n_{1p}(m_{1p}),u^n_{1c}(m_{1c}),u^n_{2c}(1,L^*_{2c}),y^n_1\right)\in A^{(n)}_{\varepsilon 1} ~~ {\rm  for}~ m_{1p} \ne 1 ~{\rm  and }~m_{1c} \ne 1 \big\}\IEEEnonumber\\
E^{dec1}_{5}=\big\{\left(q^n,u^n_{1p}(m_{1p}),u^n_{1c}(1),u^n_{2c}(m_{2c},l_{2c}),y^n_1\right)\in A^{(n)}_{\varepsilon 1} ~~ {\rm  for}~ m_{1p} \ne 1, ~ m_{2c} \ne 1 ~{\rm  and }~l_{2c} \ne L^*_{2c} \big\}\IEEEnonumber\\
E^{dec1}_{6}=\big\{\left(q^n,u^n_{1p}(1),u^n_{1c}(m_{1c}),u^n_{2c}(m_{2c},l_{2c}),y^n_1\right)\in A^{(n)}_{\varepsilon 1} ~~ {\rm  for}~ m_{1c} \ne 1, ~ m_{2c} \ne 1 ~{\rm  and }~l_{2c} \ne L^*_{2c} \big\}\IEEEnonumber\\
E^{dec1}_{7}=\big\{\left(q^n,u^n_{1p}(m_{1p}),u^n_{1c}(m_{1c}),u^n_{2c}(m_{2c},l_{2c}),y^n_1\right)\in A^{(n)}_{\varepsilon 1} ~~ {\rm  for}~ m_{1p} \ne 1, ~m_{1c} \ne 1, ~ m_{2c} \ne 1 \IEEEnonumber\\
~~~~~~~~~~~~~~~~~~~~~~~~~~~~~~~~~~~~~~~~~~~~~~~~~~~~~~~~~~~~~~~~~~~~~~~~~~~~~~~~~~~~~~~~~~~~~~{\rm  and }~l_{2c} \ne L^*_{2c} \big\}\IEEEnonumber
\label{Dec1.errors} 
\end{IEEEeqnarray} 
where, for simplicity, $A^{(n)}_{\varepsilon }\left(Q,U_{1p},U_{1c},U_{2c},Y_1\right)$ is denoted by $A^{(n)}_{\varepsilon 1}$. Note that the probability of decoding error events will be evaluated by considering: (i) the encoding distribution \eqref{pdf_DMT} (i.e., ${\mathcal P}^{DMT}$) and the actual transmitted sequences, and (ii) the codebook generation distribution \eqref{pdf_DMT_codebook}, the correctly decoded sequences and how to generate the sequences. As we mentioned earlier, for decoder 1 only the rate-terms (2.7)--(2.9) are wrong, therefore we only evaluate the probabilities of $E^{dec1}_{5}$, $E^{dec1}_{6}$ and $E^{dec1}_{7}$. The probability of the event $E^{dec1}_5$ can be bounded as
\begin{align}
&P_r\left(E^{dec1}_{5}\right) = P\Bigg\{\bigcup_{m_{1p} \ne 1 , m_{2c} \ne 1 , l_{2c}\ne L^*_{2c}}{\left(q^n,u^n_{1p}(m_{1p}),u^n_{1c}(1),u^n_{2c}(m_{2c},l_{2c}),y^n_1\right)\in A^{(n)}_{\varepsilon 1}}\Bigg\}  \IEEEnonumber\\ 
&~ \le 2^{n\left(R_{1p}+R_{2c}+R^{'}_{2c}\right)}\sum_{\left(q^n,u^n_{1p},u^n_{1c},u^n_{2c},y^n_1\right)\in A^n_{\varepsilon 1}}\bigg\{p\left(q^n\right)p\left(u^n_{1p}|q^n\right)p\left(u^n_{1c}|q^n\right)p\left(u^n_{2c}|q^n\right)p\left(y^n_1|q^n,u^n_{1c}\right)\bigg\} \IEEEnonumber\\
&~\le 2^{n\left(R_{1p}+R_{2c}+R^{'}_{2c}\right)}2^{nH\left(Q,U_{1p},U_{1c},U_{2c},Y_1\right)}2^{-nH\left(Q\right)}2^{-nH\left(U_{1p}|Q\right)}2^{-nH\left(U_{1c}|Q\right)}2^{-nH\left(U_{2c}|Q\right)}2^{-nH\left(Y_1|Q,U_{1c}\right)}  \IEEEnonumber\\
&= 2^{n\left(R_{1p}+R_{2c}+R^{'}_{2c}\right)}2^{nH\left(Q\right)}2^{nH\left(U_{1p}|Q\right)}2^{nH\left(U_{1c}|Q\right)}2^{nH\left(U_{2c}|Q,U_{1p},U_{1c}\right)}2^{nH\left(Y_1|Q,U_{1p},U_{1c},U_{2c}\right)} \IEEEnonumber\\
&~~~~~~~~~~~~~~~~~~~\times 2^{-nH\left(Q\right)}2^{-nH\left(U_{1p}|Q\right)}2^{-nH\left(U_{1c}|Q\right)}2^{-nH\left(U_{2c}|Q\right)}2^{-nH\left(Y_1|Q,U_{1c}\right)}  \IEEEnonumber\\
&~=2^{-n\big(I\left(Y_1;U_{1p},U_{2c}|Q,U_{1c}\right)+I\left(U_{2c};U_{1p},U_{1c}|Q\right)-\left(R_{1p}+R_{2c}+R^{'}_{2c}\right)\big)}  
\label{Dec1.error5.eval}
\end{align}
Hence, $P_r\left(E^{dec1}_5\right)$ goes to zero as $n\to \infty $ if (3.7) is satisfied. Similarly, $P_r\left(E^{dec1}_{6}\right)$ and $P_r\left(E^{dec1}_{7}\right)$ can be bounded as
\begin{align}
&P_r\left(E^{dec1}_{6}\right) = P\Bigg\{\bigcup_{m_{1c} \ne 1 , m_{2c} \ne 1 , l_{2c}\ne L^*_{2c}}{\left(q^n,u^n_{1p}(1),u^n_{1c}(m_{1c}),u^n_{2c}(m_{2c},l_{2c}),y^n_1\right)\in A^{(n)}_{\varepsilon 1}}\Bigg\}  \IEEEnonumber\\ 
&~ \le 2^{n\left(R_{1c}+R_{2c}+R^{'}_{2c}\right)}\sum_{\left(q^n,u^n_{1p},u^n_{1c},u^n_{2c},y^n_1\right)\in A^n_{\varepsilon 1}}\bigg\{p\left(q^n\right)p\left(u^n_{1p}|q^n\right)p\left(u^n_{1c}|q^n\right)p\left(u^n_{2c}|q^n\right)p\left(y^n_1|q^n,u^n_{1p}\right)\bigg\} \IEEEnonumber\\
&~\le 2^{n\left(R_{1c}+R_{2c}+R^{'}_{2c}\right)}2^{nH\left(Q,U_{1p},U_{1c},U_{2c},Y_1\right)}2^{-nH\left(Q\right)}2^{-nH\left(U_{1p}|Q\right)}2^{-nH\left(U_{1c}|Q\right)}2^{-nH\left(U_{2c}|Q\right)}2^{-nH\left(Y_1|Q,U_{1p}\right)}  \IEEEnonumber\\
&~=2^{-n\big(I\left(Y_1;U_{1c},U_{2c}|Q,U_{1p}\right)+I\left(U_{2c};U_{1p},U_{1c}|Q\right)-\left(R_{1c}+R_{2c}+R^{'}_{2c}\right)\big)}  
\label{Dec1.error6.eval}
\end{align}
\begin{align}
&P_r\left(E^{dec1}_{7}\right) = P\Bigg\{\bigcup_{m_{1p} \ne 1 , m_{1c} \ne 1 , m_{2c} \ne 1 , l_{2c}\ne L^*_{2c}}{\left(q^n,u^n_{1p}(m_{1p}),u^n_{1c}(m_{1c}),u^n_{2c}(m_{2c},l_{2c}),y^n_1\right)\in A^{(n)}_{\varepsilon 1}}\Bigg\}  \IEEEnonumber\\ 
&~ \le 2^{n\left(R_{1p}+R_{1c}+R_{2c}+R^{'}_{2c}\right)}\sum_{\left(q^n,u^n_{1p},u^n_{1c},u^n_{2c},y^n_1\right)\in A^n_{\varepsilon 1}}\bigg\{p\left(q^n\right)p\left(u^n_{1p}|q^n\right)p\left(u^n_{1c}|q^n\right)p\left(u^n_{2c}|q^n\right)p\left(y^n_1|q^n\right)\bigg\} \IEEEnonumber\\
&~\le 2^{n\left(R_{1p}+R_{1c}+R_{2c}+R^{'}_{2c}\right)}2^{nH\left(Q,U_{1p},U_{1c},U_{2c},Y_1\right)}2^{-nH\left(Q\right)}2^{-nH\left(U_{1p}|Q\right)}2^{-nH\left(U_{1c}|Q\right)}2^{-nH\left(U_{2c}|Q\right)}2^{-nH\left(Y_1|Q\right)}  \IEEEnonumber\\
&~=2^{-n\big(I\left(Y_1;U_{1p},U_{1c},U_{2c}|Q\right)+I\left(U_{2c};U_{1p},U_{1c}|Q\right)-\left(R_{1p}+R_{1c}+R_{2c}+R^{'}_{2c}\right)\big)}  
\label{Dec1.error7.eval}
\end{align}
Hence, $P_r\left(E^{dec1}_{6}\right)\to 0 $ and $P_r\left(E^{dec1}_{7}\right)\to 0 $ as $n\to \infty $ if (3.8) and (3.9) are satisfied, respectively. 

The determining error events at decoder 2 are  
\begin{IEEEeqnarray}{l}
E^{dec2}_{1}=\big\{\left(q^n,u^n_{2p}(m_{2p},l_{2p}),u^n_{2c}(1,L^*_{2c}),u^n_{1c}(1),y^n_2\right)\in A^{(n)}_{\varepsilon 2} ~~ {\rm  for}~ m_{2p} \ne 1 ~{\rm  and }~l_{2p} \ne L^*_{2p} \big\}\IEEEnonumber\\
E^{dec2}_{2}=\big\{\left(q^n,u^n_{2p}(1,L^*_{2p}),u^n_{2c}(m_{2c},l_{2c}),u^n_{1c}(1),y^n_2\right)\in A^{(n)}_{\varepsilon 2} ~~ {\rm  for}~ m_{2c} \ne 1 ~{\rm  and }~l_{2c} \ne L^*_{2c} \big\}\IEEEnonumber\\
E^{dec2}_{3}=\big\{\left(q^n,u^n_{2p}(1,L^*_{2p}),u^n_{2c}(1,L^*_{2c}),u^n_{1c}(m_{1c}),y^n_2\right)\in A^{(n)}_{\varepsilon 2} ~~ {\rm  for}~ m_{1c} \ne 1  \big\}\IEEEnonumber\\
E^{dec2}_{4}=\big\{\left(q^n,u^n_{2p}(m_{2p},l_{2p}),u^n_{2c}(m_{2c},l_{2c}),u^n_{1c}(1),y^n_2\right)\in A^{(n)}_{\varepsilon 2} ~~ {\rm  for}~ m_{2p} \ne 1,~l_{2p} \ne L^*_{2p}, m_{2c} \ne 1 \IEEEnonumber\\
~~~~~~~~~~~~~~~~~~~~~~~~~~~~~~~~~~~~~~~~~~~~~~~~~~~~~~~~~~~~~~~~~~~~~~~~~~~~~~~~~~~~~~~~~~~~~~~{\rm  and }~l_{2c} \ne L^*_{2c} \big\}\IEEEnonumber\\
E^{dec2}_{5}=\big\{\left(q^n,u^n_{2p}(m_{2p},l_{2p}),u^n_{2c}(1,L^*_{2c}),u^n_{1c}(m_{1c}),y^n_2\right)\in A^{(n)}_{\varepsilon 2} ~~ {\rm  for}~m_{2p} \ne 1,~l_{2p} \ne L^*_{2p},~ m_{1c} \ne 1  \big\}\IEEEnonumber\\
E^{dec2}_{6}=\big\{\left(q^n,u^n_{2p}(1,L^*_{2p}),u^n_{2c}(m_{2c},l_{2c}),u^n_{1c}(m_{1c}),y^n_2\right)\in A^{(n)}_{\varepsilon 2} ~~ {\rm  for}~ m_{2c} \ne 1, ~l_{2c} \ne L^*_{2c},~m_{1c} \ne 1 \big\}\IEEEnonumber\\
E^{dec2}_{7}=\big\{\left(q^n,u^n_{2p}(m_{2p},l_{2p}),u^n_{2c}(m_{2c},l_{2c}),u^n_{1c}(m_{1c}),y^n_2\right)\in A^{(n)}_{\varepsilon 2} ~~ {\rm  for}~ m_{2p} \ne 1,~l_{2p} \ne L^*_{2p}, m_{2c} \ne 1, \IEEEnonumber\\
~~~~~~~~~~~~~~~~~~~~~~~~~~~~~~~~~~~~~~~~~~~~~~~~~~~~~~~~~~~~~~~~~~~~~~~~~~~~~~~~~~~~l_{2c} \ne L^*_{2c}~ {\rm  and }~m_{1c} \ne 1 \big\} \IEEEnonumber
\label{Dec2.errors} 
\end{IEEEeqnarray} 
where, for simplicity, $A^{(n)}_{\varepsilon }\left(Q,U_{2p},U_{2c},U_{1c},Y_2\right)$ is denoted by $A^{(n)}_{\varepsilon 2}$. As we mentioned earlier, for decoder 2 only the rate-terms (2.13)--(2.16) are wrong, therefore we only evaluate the probabilities of $E^{dec2}_{4}$, $E^{dec2}_{5}$, $E^{dec2}_{6}$ and $E^{dec2}_{7}$. The probability of the event $E^{dec2}_4$ can be bounded as

\begin{align}
&P_r\left(E^{dec2}_{4}\right) = P\Bigg\{\bigcup_{m_{2p} \ne 1, l_{2p} \ne L^*_{2p}, m_{2c} \ne 1, l_{2c} \ne L^*_{2c}}{\left(q^n,u^n_{2p}(m_{2p},l_{2p}),u^n_{2c}(m_{2c},l_{2c}),u^n_{1c}(1),y^n_2\right)\in A^{(n)}_{\varepsilon 2}}\Bigg\}  \IEEEnonumber\\ 
&~ \le 2^{n\left(R_{2p}+R^{'}_{2p}+R_{2c}+R^{'}_{2c}\right)}\sum_{\left(q^n,u^n_{2p},u^n_{2c},u^n_{1c},y^n_2\right)\in A^n_{\varepsilon 2}}\bigg\{p\left(q^n\right)p\left(u^n_{2p}|q^n\right)p\left(u^n_{2c}|q^n\right)p\left(u^n_{1c}|q^n\right)p\left(y^n_2|q^n,u^n_{1c}\right)\bigg\} \IEEEnonumber\\
&~\le 2^{n\left(R_{2p}+R^{'}_{2p}+R_{2c}+R^{'}_{2c}\right)}2^{nH\left(Q,U_{2p},U_{2c},U_{1c},Y_2\right)}2^{-nH\left(Q\right)}2^{-nH\left(U_{2p}|Q\right)}2^{-nH\left(U_{2c}|Q\right)}2^{-nH\left(U_{1c}|Q\right)}2^{-nH\left(Y_2|Q,U_{1c}\right)}  \IEEEnonumber\\
&~=2^{-n\big(I\left(Y_2;U_{2p},U_{2c}|Q,U_{1c}\right)+I\left(U_{2p};U_{2c}|Q\right)+I\left(U_{2p},U_{2c};U_{1c}|Q\right)-\left(R_{2p}+R^{'}_{2p}+R_{2c}+R^{'}_{2c}\right)\big)}  
\label{Dec2.error4.eval}
\end{align}
where, \eqref{Dec2.error4.eval} is obtained by considering this fact $U_{2p}$ and $U_{2c}$ are dependent in general and the encoding distribution \eqref{pdf_DMT} says that they are independent only when $(U_{1p},U_{1c},Q)$ are given, i.e., $U_{2p} \rightarrow (U_{1p},U_{1c},Q) \rightarrow U_{2c}$ form a Markov chain. Hence, $P_r\left(E^{dec2}_4\right)$ goes to zero as $n\to \infty $ if (3.13) is satisfied. Similarly, $P_r\left(E^{dec2}_{5}\right)$, $P_r\left(E^{dec2}_{6}\right)$ and $P_r\left(E^{dec1}_{7}\right)$ can be bounded as
\begin{align}
&P_r\left(E^{dec2}_{5}\right) = P\Bigg\{\bigcup_{m_{2p} \ne 1, l_{2p} \ne L^*_{2p}, m_{1c} \ne 1}{\left(q^n,u^n_{2p}(m_{2p},l_{2p}),u^n_{2c}(1,L^*_{2c}),u^n_{1c}(m_{1c}),y^n_2\right)\in A^{(n)}_{\varepsilon 2}}\Bigg\}  \IEEEnonumber\\ 
&~ \le 2^{n\left(R_{2p}+R^{'}_{2p}+R_{1c}\right)}\sum_{\left(q^n,u^n_{2p},u^n_{2c},u^n_{1c},y^n_2\right)\in A^n_{\varepsilon 2}}\bigg\{p\left(q^n\right)p\left(u^n_{2p}|q^n\right)p\left(u^n_{2c}|q^n\right)p\left(u^n_{1c}|q^n\right)p\left(y^n_2|q^n,u^n_{2c}\right)\bigg\} \IEEEnonumber\\
&~\le 2^{n\left(R_{2p}+R^{'}_{2p}+R_{1c}\right)}2^{nH\left(Q,U_{2p},U_{2c},U_{1c},Y_2\right)}2^{-nH\left(Q\right)}2^{-nH\left(U_{2p}|Q\right)}2^{-nH\left(U_{2c}|Q\right)}2^{-nH\left(U_{1c}|Q\right)}2^{-nH\left(Y_2|Q,U_{2c}\right)}  \IEEEnonumber\\
&~=2^{-n\big(I\left(Y_2;U_{2p},U_{1c}|Q,U_{2c}\right)+I\left(U_{2p};U_{1c}|Q\right)+I\left(U_{2c};U_{2p},U_{1c}|Q\right)-\left(R_{2p}+R^{'}_{2p}+R_{1c}\right)\big)}  
\label{Dec2.error5.eval}
\end{align}
\begin{align}
&P_r\left(E^{dec2}_{6}\right) = P\Bigg\{\bigcup_{m_{2c} \ne 1, l_{2c} \ne L^*_{2c}, m_{1c} \ne 1}{\left(q^n,u^n_{2p}(1,L^*_{2p}),u^n_{2c}(m_{2c},l_{2c}),u^n_{1c}(m_{1c}),y^n_2\right)\in A^{(n)}_{\varepsilon 2}}\Bigg\}  \IEEEnonumber\\ 
&~ \le 2^{n\left(R_{2c}+R^{'}_{2c}+R_{1c}\right)}\sum_{\left(q^n,u^n_{2p},u^n_{2c},u^n_{1c},y^n_2\right)\in A^n_{\varepsilon 2}}\bigg\{p\left(q^n\right)p\left(u^n_{2p}|q^n\right)p\left(u^n_{2c}|q^n\right)p\left(u^n_{1c}|q^n\right)p\left(y^n_2|q^n,u^n_{2p}\right)\bigg\} \IEEEnonumber\\
&~\le 2^{n\left(R_{2c}+R^{'}_{2c}+R_{1c}\right)}2^{nH\left(Q,U_{2p},U_{2c},U_{1c},Y_2\right)}2^{-nH\left(Q\right)}2^{-nH\left(U_{2p}|Q\right)}2^{-nH\left(U_{2c}|Q\right)}2^{-nH\left(U_{1c}|Q\right)}2^{-nH\left(Y_2|Q,U_{2p}\right)}  \IEEEnonumber\\
&~=2^{-n\big(I\left(Y_2;U_{2c},U_{1c}|Q,U_{2p}\right)+I\left(U_{2p};U_{2c},U_{1c}|Q\right)+I\left(U_{2c};U_{1c}|Q\right)-\left(R_{2c}+R^{'}_{2c}+R_{1c}\right)\big)}  
\label{Dec2.error6.eval}
\end{align}
\begin{align}
&P_r\left(E^{dec2}_{7}\right) = P\Bigg\{\bigcup_{m_{2p} \ne 1, l_{2p} \ne L^*_{2p}, m_{2c} \ne 1, l_{2c} \ne L^*_{2c}, m_{1c} \ne 1}{\left(q^n,u^n_{2p}(m_{2p},l_{2p}),u^n_{2c}(m_{2c},l_{2c}),u^n_{1c}(m_{1c}),y^n_2\right)\in A^{(n)}_{\varepsilon 2}}\Bigg\}  \IEEEnonumber\\ 
&~ \le 2^{n\left(R_{2p}+R^{'}_{2p}+R_{2c}+R^{'}_{2c}+R_{1c}\right)}\sum_{\left(q^n,u^n_{2p},u^n_{2c},u^n_{1c},y^n_2\right)\in A^n_{\varepsilon 2}}\bigg\{p\left(q^n\right)p\left(u^n_{2p}|q^n\right)p\left(u^n_{2c}|q^n\right)p\left(u^n_{1c}|q^n\right)p\left(y^n_2|q^n\right)\bigg\} \IEEEnonumber\\
&~\le 2^{n\left(R_{2p}+R^{'}_{2p}+R_{2c}+R^{'}_{2c}+R_{1c}\right)}2^{nH\left(Q,U_{2p},U_{2c},U_{1c},Y_2\right)}2^{-nH\left(Q\right)}2^{-nH\left(U_{2p}|Q\right)}2^{-nH\left(U_{2c}|Q\right)}2^{-nH\left(U_{1c}|Q\right)}2^{-nH\left(Y_2|Q\right)}  \IEEEnonumber\\
&~=2^{-n\big(I\left(Y_2;U_{2p},U_{2c},U_{1c}|Q\right)+I\left(U_{2p};U_{2c}|Q\right)+I\left(U_{2p},U_{2c};U_{1c}|Q\right)-\left(R_{2p}+R^{'}_{2p}+R_{2c}+R^{'}_{2c}+R_{1c}\right)\big)}  
\label{Dec2.error7.eval}
\end{align}
Hence, $P_r\left(E^{dec2}_{5}\right)\to 0 $, $P_r\left(E^{dec2}_{6}\right)\to 0 $ and $P_r\left(E^{dec2}_{7}\right)\to 0 $ as $n\to \infty $ if (3.14), (3.15) and (3.9) are satisfied, respectively. This completes the proof of Theorem 2.


\end{document}